**Hyperbolic metamaterials: fundamentals and applications**
Prashant Shekhar, Jonathan Atkinson and Zubin Jacob*

Department of Electrical and Computer Engineering,
University of Alberta, Edmonton, Canada, AB T6G 2V4
emails: pshekhar@ualberta.ca, jcatkins@ualberta.ca,

Corresponding author: zjacob@ualberta.ca

**Abstract:** Metamaterials are nano-engineered media with designed properties beyond those available in nature with applications in all aspects of materials science. In particular, metamaterials have shown promise for next generation optical materials with electromagnetic responses that cannot be obtained from conventional media. We review the fundamental properties of metamaterials with hyperbolic dispersion and present the various applications where such media offer potential for transformative impact. These artificial materials support unique bulk electromagnetic states which can tailor light-matter interaction at the nanoscale. We present a unified view of practical approaches to achieve hyperbolic dispersion using thin film and nanowire structures. We also review current research in the field of hyperbolic metamaterials such as sub-wavelength imaging and broadband photonic density of states engineering. The review introduces the concepts central to the theory of hyperbolic media as well as nanofabrication and characterization details essential to experimentalists. Finally, we outline the challenges in the area and offer a set of directions for future work.

## 1.0 Introduction

Metamaterials research has captured the imagination of optical engineers and materials scientists because of their varied applications including imaging [1]–[3], cloaking [4], [5], sensing [6], waveguiding [7], and simulating space-time phenomena [8]. The field of metamaterials started with the search for negative dielectric permittivity and magnetic permeability [9] however the range of electromagnetic responses achievable using nanostructured media far surpass the concept of negative index. The invisibility cloak is the best example where an inhomogeneous anisotropic electromagnetic response causes light to bend smoothly around an object rendering it invisible [10]. Another example is that of chiral metamaterials, where the response of a medium to polarized light can be enhanced by orders of magnitude through artificial structures [11], [12].

While all the above media have specific domains of application, hyperbolic metamaterials are a multi-functional platform to realize waveguiding, imaging, sensing, quantum and thermal engineering beyond conventional devices [13]–[18]. This metamaterial uses the concept of engineering the basic dispersion relation of waves to provide unique electromagnetic modes that can have a broad range of applications [19], [20]. One can consider the hyperbolic metamaterial as a polaritonic crystal where the coupled states of light and matter give rise to a larger bulk density of electromagnetic states [21], [22]. Some of the applications of hyperbolic metamaterials include negative refraction [23], [24], sub-diffraction imaging [3], [25], sub-wavelength modes [7], [26], and spontaneous [27]–[31] and thermal emission engineering [32], [33] [34].

The initial work in artificial structures with hyperbolic behavior started in the microwave regime (indefinite media) with phenomena such as resonance cones [18], negative refraction [35] and canalization of images [15]. In the optical domain, it was proposed that non-magnetic media can show hyperbolic behavior leading to negative index waveguides [14], sub-wavelength imaging [3] and sub-diffraction photonic funnels [7]. This review aims to provide an overview of the properties of hyperbolic media from an experimental perspective focusing on design and characterization [36]. To aid experimentalists, we describe the different experimental realizations of hyperbolic media (thin film and nanowire geometry) and contrast their differences [23], [24]. After initial sections on design and characterization, we also review applications of sub-diffraction imaging and density of states engineering with hyperbolic metamaterials. We give a detailed section for experimentalists to analyze the various figures of merit related to hyperbolic media and spontaneous emission engineering.

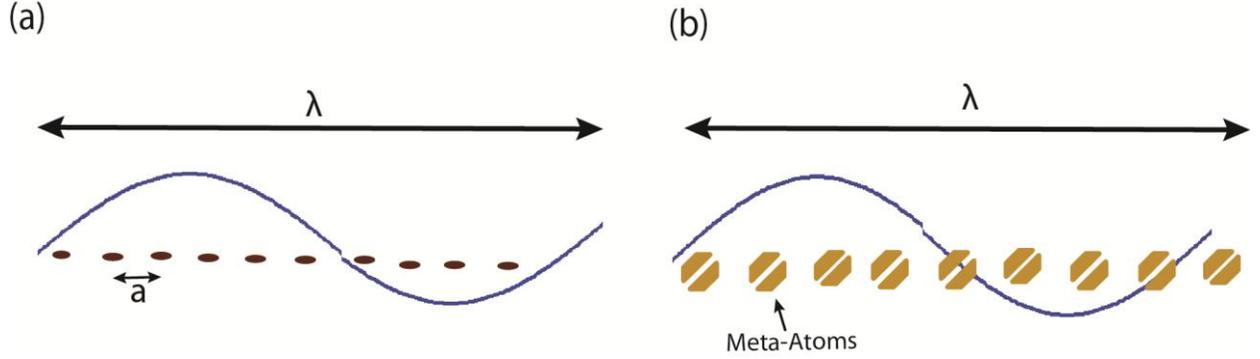

Figure 1: (a) A periodic array of atoms with the radius and inter-atomic spacing of each atom being much less than the wavelength of radiation. (b) Metamaterials are composed of nano-inclusions called meta atoms with critical dimensions much less than the wavelength of radiation. The artificial meta atoms provide unique electromagnetic responses not seen with natural structures.

Hyperbolic metamaterials (HMMs) derive their name from the topology of the isofrequency surface. In vacuum, the linear dispersion and isotropic behavior of propagating waves implies a spherical isofrequency surface given by the equation $k_x^2 + k_y^2 + k_z^2 = \omega^2 / c^2$ (figure 2(a)). Here, the wavevector of a propagating wave is given by $\vec{k} = [k_x, k_y, k_z]$, $\omega$ is the frequency of radiation and $c$ is the velocity of light in free space. If we consider an extraordinary wave (TM polarized) in a uniaxial medium, this isofrequency relation changes to

$$\frac{k_x^2 + k_y^2}{\varepsilon_{zz}} + \frac{k_z^2}{\varepsilon_{xx}} = \frac{\omega^2}{c^2} \quad (1)$$

Note the uniaxial medium has a dielectric response given by a tensor $\vec{\varepsilon} = \left[ \varepsilon_{xx}, \varepsilon_{yy}, \varepsilon_{zz} \right]$ where the in-plane isotropic components are $\varepsilon_{xx} = \varepsilon_{yy} = \varepsilon_{\parallel}$ and out of plane component is $\varepsilon_{zz} = \varepsilon_{\perp}$. The spherical isofrequency surface of vacuum distorts to an ellipsoid for the anisotropic case.

However, when we have extreme anisotropy such that $\varepsilon_{\parallel} \cdot \varepsilon_{\perp} < 0$, the isofrequency surface opens into an open hyperboloid (figure 2(b,c)). Such a phenomenon requires the material to behave like a metal in one direction and a dielectric (insulator) in the other. This does not readily occur in nature at optical frequencies but can be achieved using artificial nanostructured electromagnetic media: metamaterials.

The most important property of such media is related to the behavior of waves with large magnitude wavevectors. In vacuum, such large wavevector waves are evanescent and decay exponentially. However, in hyperbolic media the open form of the isofrequency surface allows for propagating waves with infinitely large wavevectors in the idealistic limit [3], [27]. Thus there are no evanescent waves in such a medium. This unique property of propagating high-$k$ waves gives rise to a multitude of device applications using hyperbolic media [19], [20].

We introduce a classification for hyperbolic media that helps to identify their properties. Type I HMMs have one component of the dielectric tensor negative ($\varepsilon_{zz} < 0$; $\varepsilon_{xx}$; $\varepsilon_{yy} > 0$) while Type II HMMs have two components negative ($\varepsilon_{xx}$; $\varepsilon_{yy} < 0$; $\varepsilon_{zz} > 0$) and are shown in figure 2(b) and figure 2(c), respectively. Note of course, that if all components are negative, we obtain a metal and if all components are positive we will have a dielectric medium. One striking difference between the Type I and Type II hyperbolic metamaterial is that the hyperboloidal surfaces are two sheeted and single sheeted respectively. The Type II metamaterial is highly reflective since it is more metallic than the Type I counterpart [37].

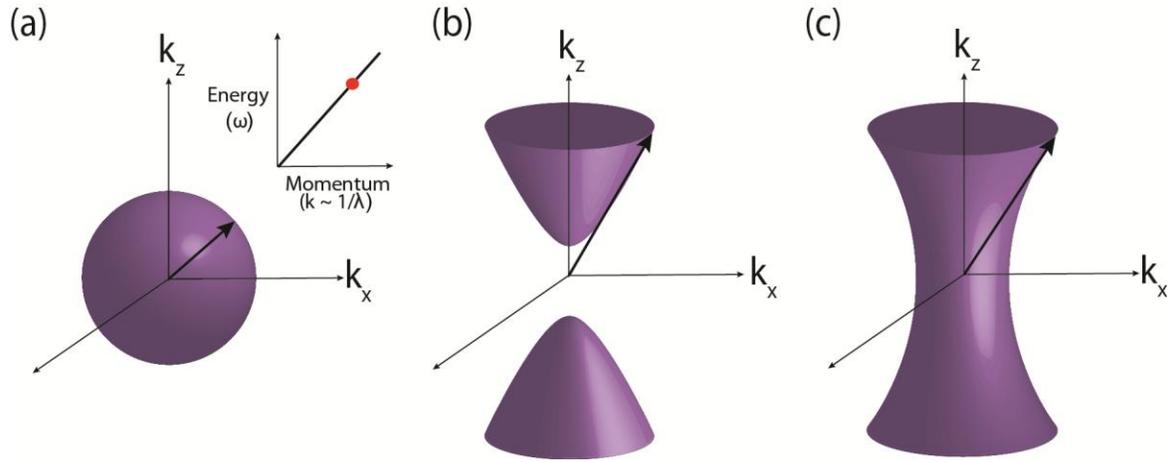

Figure 2: k-space topology. (a) Spherical isofrequency surface for an isotropic dielectric. Inset shows an energy versus momentum relationship with the red dot indicating the operating frequency for the derived isofrequency surface. (b) Hyperboloid isofrequency surface for a uniaxial medium with an extremely anistropic dielectric response (Type I HMM: $\varepsilon_{zz} < 0$; $\varepsilon_{xx}$; $\varepsilon_{yy} > 0$) (c) Hyperboloid isofrequency surface for an extremely anistropic uniaxial medium with two negative components of the dielectric tensor (Type II HMM: $\varepsilon_{xx}$; $\varepsilon_{yy} < 0$; $\varepsilon_{zz} > 0$). The (b) Type I and (c) Type II metamaterials can support waves with infinitely large wavevectors in the effective medium limit. Such waves are evanescent and decay away exponentially in vacuum.

## 2.0 Design and Materials

There are two practical approaches to achieve the hyperbolic dispersion which we discuss below. The fact that hyperbolicity requires metallic behavior in one direction and insulating behavior in the other leads to the requirement that both metals and dielectrics must be used as building blocks. Microscopically, the origin of the high-k propagating waves relies on a metallic building block to create the hyperbolic dispersion of the material (section 1.0). The polaritonic properties of the metallic building blocks allow for the necessary light-matter coupling to create the high-k waves. Specifically, it is necessary to have a phonon-polaritonic (optically active phonons) or plasmon-polaritonic (free electron) metal to construct hyperbolic metamaterials. The high-k modes are a result of the near-field coupling of the surface plasmon polaritons (SPPs) at each of the metal-dielectric interfaces in the structure. The high-k modes are the Bloch modes of the metal-dielectric superlattice.

a) 1D HMM

A thin film multilayer (super-lattice) consisting of alternating layers of metal and dielectric gives rise to the desired extreme anisotropy [38]. The layer thicknesses should be far below the size of the operating wavelength for the homogenization to be valid. A detailed derivation is given in appendix 1.0. We emphasize right at the outset that the most important figure of merit for the HMM is the plasma frequency of the metal and the loss. These two quantities determine the impedance matching and absorption of the metamaterial for practical applications [39].

A wide choice of plasmonic metals and high index dielectrics can give rise to hyperbolic behavior in different wavelength regimes. At ultraviolet (UV) frequencies, gold and silver along with alumina forms the ideal choice for the metamaterial. Close to the plasma frequency of these metals, which is in the UV, their reflectivity decreases and an alternating metal dielectric super-lattice achieves a Type I HMM with high transmission. To push this design to visible wavelengths however a high index dielectric such as $TiO_2$ or SiN is needed [40].

At near-infrared (IR) wavelengths, compensating for the reflective metallic behavior of naturally plasmonic metals like silver and gold is unfeasible and alternate plasmonic materials with tailored lower plasma frequencies are needed. These alternate plasmonic materials are based on transition metal nitrides or transparent conducting oxides and are ideally suited for hyperbolic media [39], [41]. Recently, their unique property of high melting point was also used to pave the way for high temperature thermal hyperbolic metamaterials [42], [43].

At mid-infrared wavelengths, one option for the metallic component in hyperbolic media consists of III-V degenerately doped semiconductors [24], [44]. The upper limit of doping concentration often limits their abilities to work as a metal at near-IR wavelengths, however they are ideally suited to the mid-IR. Another option which is fundamentally different from above mentioned plasmonic metals is silicon carbide, a low loss phonon polaritonic metal [37], [45], [46]. SiC has a narrow reststrahlen band at mid-IR wavelengths which allows it to function as a metallic building block for hyperbolic media. SiC based hyperbolic media were recently predicted to show super-Planckian thermal emission [32], [33]. Multilayer graphene super-lattices can also show a hyperbolic metamaterial response in the THz (far-IR) region of the spectrum [47]–[50].

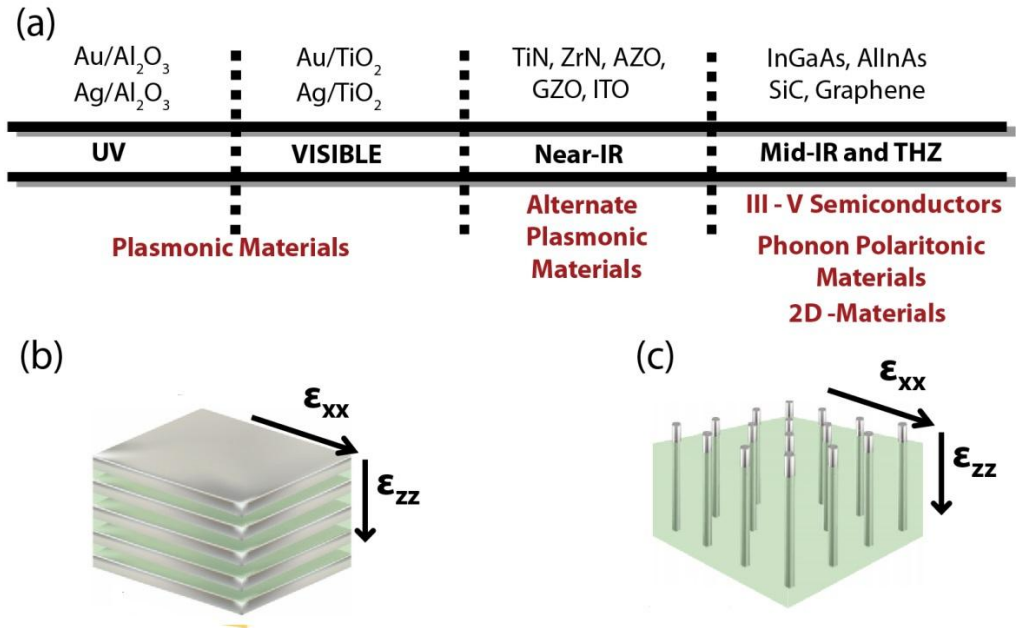

Figure 3: (a) Materials used to create hyperbolic metamaterials depending on region of operation in the electromagnetic spectrum (UV to mid-IR and THZ frequencies) (b) Multilayer structure consisting of alternating metallic and dielectric layers forming a metal-dielectric superlattice. (c) Nanowire structure consisting of metallic nanorods embedded in a dielectric host. In both (b) and (c) the constituent components are subwavelength allowing characterization with effective medium theory.

## b) 2D HMM

Another approach to achieving hyperbolic behavior consists of metallic nanowires in a dielectric host [23], [36], [51]–[53]. The choice of metals are usually silver and gold grown in a nanoporous alumina template. The major advantage of this design is the low losses, broad bandwidth and high transmission. Also, the problem of large reflectivity like the multilayer design does not exist and we can achieve Type I hyperbolic behavior. Note the fill fraction of metal needed in the 2D design to achieve Type I hyperbolic behavior is far below that in the multilayer design leading to a large figure of merit.

**3.0 Nanofabrication**

Here we give a brief overview of the popular techniques used to fabricate hyperbolic metamaterials.

a) Thin film

The multilayer design relies extensively on the deposition of ultrathin and smooth thin films of metal and dielectric. Surface roughness is surely an issue for practical applications due to increased material loss and light scattering. However, minor deviations in layer thicknesses do not appreciably change the effective medium response [54]–[56]. Gold and silver along with alumina or titanium dioxide have been deposited by multiple groups using electron beam evaporation [22], [57]. Typical layer thicknesses can include 22 nm Ag alternated with 40 nm $TiO_2$ layers. The alternate plasmonic materials can be made via reactive sputtering or deposited by pulsed laser deposition to maintain the required strict stoichiometry [39]. Silicon carbide can be grown by plasma enhanced chemical vapor deposition (PECVD) [37]. However, it is difficult to observe crystalline behavior crucial for high quality phonon-polariton resonances in multilayer structures. For III-V semiconductors, molecular beam epitaxy (MBE) is the ideal method to grow the alternating layers consisting of highly doped semiconductors behaving as a metal [24]. Extremely uniform and smooth surfaces are possible with MBE.

b) Nanowire

The standard procedure consists of either buying off the shelf anodic alumina membrane [36] or anodizing aluminum to grow the required template [6][58]. Multiple groups have successfully fabricated hyperbolic metamaterials using both of these approaches. This template forms the basic dielectric host medium with a periodic nanoporous structure into which the silver (or gold) nanowires can be electrodeposited. Typical dimensions for recent gold nanowire structures include 20-700 nm long wires with a 10-50 nm rod diameter with 40-70 nm rod separation [6]. Note the porosity controls the fill fraction of the metal and hence the dispersion of hyperbolic behavior. A multi-step controlled electrodeposition is necessary to ensure that the silver filling is consistent across the sample. Furthermore, a significant issue of silver overfilling or discontinuous islands within the pore has to be addressed after fabrication and presents a challenge to successful nanowire HMM fabrication.

## 4.0 Theoretical Characterization

We begin this section with a note that characterization of artificial media and effective medium parameter retrieval has been a controversial topic in the field of metamaterials [59]. This is primarily because the unit cells are often not subwavelength and the structures are two dimensional where an effective permittivity cannot be defined in the strict sense [60]. We emphasize that hyperbolic media suffer from neither of these drawbacks. The unit cells in both the 1D and 2D design are always deeply subwavelength (typically 20-70 nm) for propagating states, however, at very large wavevector values, non-local theory must be taken into consideration [61]. The structures are also three dimensional where the effective medium parameters show extreme anisotropy for p-polarized light. We now describe the experimental characteristics that can be used to confirm the presence of hyperbolic behavior in the samples.

a) Epsilon-near-zero and epsilon-near-pole responses

An interesting characteristic of multilayer and nanowire structures are the existence of poles and zeros in the effective medium dielectric constants. This results in an ideal method to first characterize the resonant responses and subsequently infer the hyperbolic characteristics. At these specific wavelengths a component of the dielectric tensor of the metamaterial either passes through zero (epsilon-near-zero, ENZ) [62], [63] or has a resonant pole (epsilon-near-pole, ENP) [42].

In figure 4, we plot the effective medium constants for the multilayer and nanowire structures using the homogenization formulae derived in appendix 1.0 and appendix 2.0, respectively. The multilayer sample shows an epsilon-near-zero effect as well as epsilon-near-pole resonance. Only the real parts are shown for clarity and the imaginary parts can be calculated similarly.  The multilayer consisting of alternating layers of silver and TiO$_2$ with a metallic fill fraction of 35 % shows both Type I and Type II hyperbolic behavior. The nanowire effective medium theory (EMT) parameters are shown in figure 4(b). Type I behavior, which is difficult to achieve with multilayer structures, is observed. It also shows the characteristic ENP and ENZ resonances.

The most important aspect to note about the ENZ and ENP resonances are the directions in which they occur for multilayer and nanowire samples. This fundamentally changes the reflection and transmission spectrum of the two types of hyperbolic media. For the two designs, ENZ occurs parallel to the thin film layers or along the nanowire length. This is intuitively expected since the Drude plasma frequency which determines the ENZ always occurs in the direction of free electron motion. Conversely, the resonant ENP behavior of the two geometries occurs in the direction for which there is no continuous free electron motion. The ENP resonance occurs perpendicular to the thin film layers in the multilayer structure and perpendicular to the wires in the nanowire geometry [42]. The directions of ENZ and ENP behavior for the multilayer and nanowire structures are shown in the schematic insets of figure 4.

b) Propagating wave spectrum

Characterization of the metamaterials is most easily done by analyzing the reflection and transmission spectrum of propagating waves. Note that measuring phase is difficult hence it is preferred to study the angle resolved reflection and transmission spectrum to infer the effective medium parameters. The features of hyperbolic behavior are manifested only in p-polarized light so it is best to study the transverse electric (s-polarized) and p-polarized light separately and contrast the differences [22], [36]. Note, that in the following analysis of the reflection, transmission, and extinction spectra, the structures do differ slightly to highlight key features. Specifics of the structures used for the analysis are highlighted in the figure captions.

*Reflection spectrum*: The Type I metamaterials have only one component of the dielectric tensor negative and are less reflective due to no free electron motion parallel to the interface. They have properties of conventional dielectrics, such as the Brewster angle, which can be used to extract the EMT parameters. Both multilayer and nanowire samples show Type I behavior as can be seen in figure 5. The Type II metamaterials have two components of the dielectric tensor negative and are highly reflective at all angles (figure 5(a)). They have properties common to conventional metals such as surface plasmon polaritons. It is easy in multilayer structures to obtain Type II behavior while it is easier in nanowire samples to obtain Type I behavior.

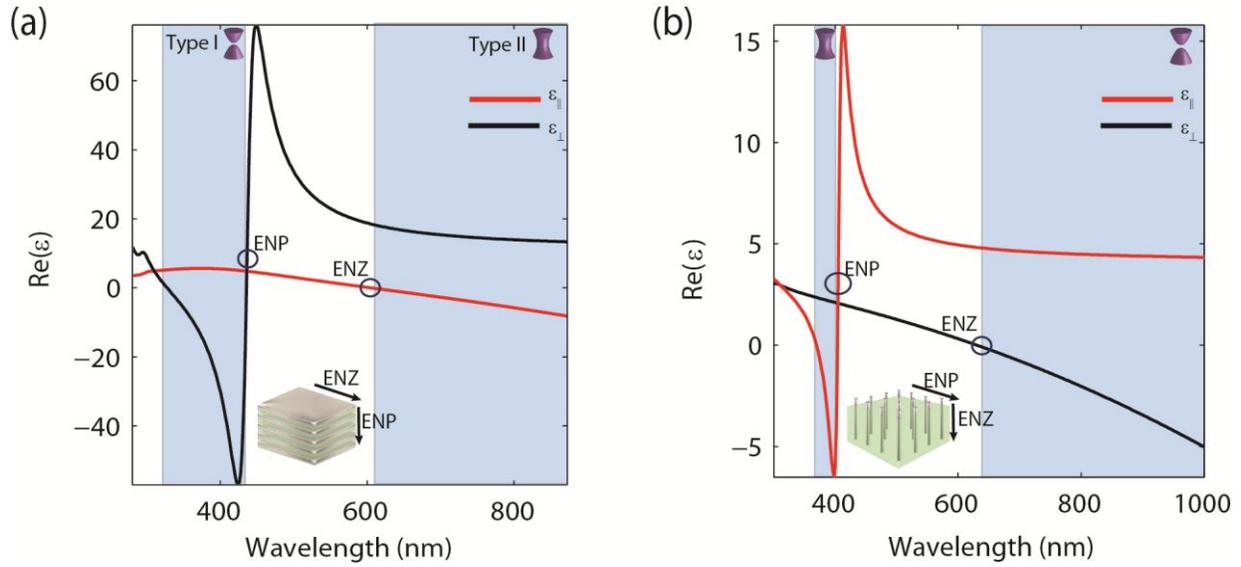

Figure 4: (a) Multilayer System: Real part of the dielectric permittivity for an Ag-TiO$_2$ multilayer structure using 35% silver fraction and effective medium theory. (b) Nanowire System: Real part of the dielectric permittivity for an Ag-Al$_2$O$_3$ nanowire structure at 15 % silver fill fraction. The Type I and Type II hyperbolic regions of the dispersion and regions of epsilon near zero (ENZ) and epsilon near pole (ENP) are highlighted. The schematic insets in both (a) and (b) show the directions of ENZ and ENP behavior.

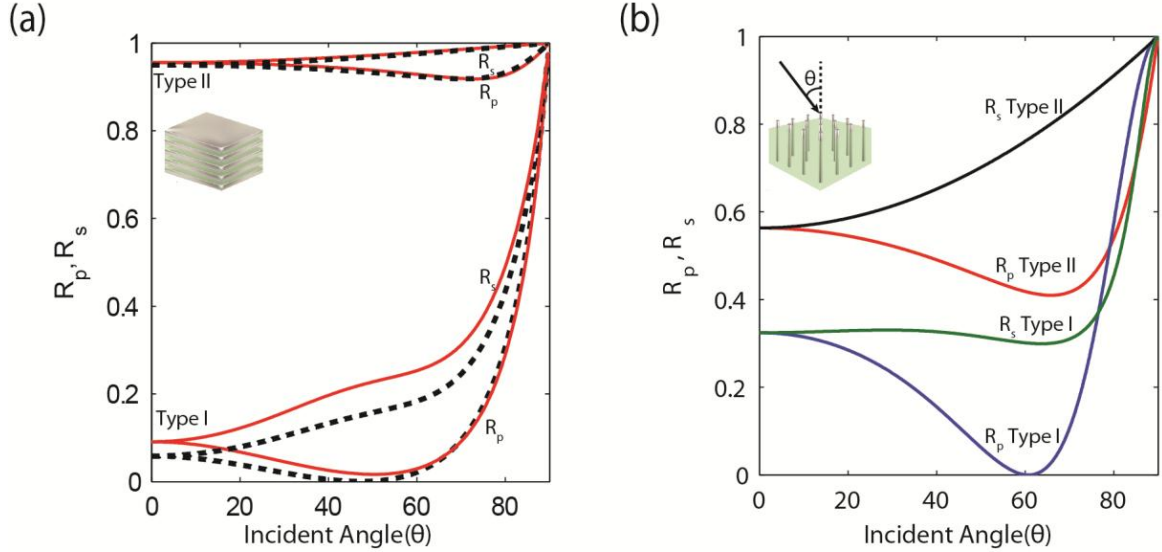

Figure 5: (a) Ag-TiO$_2$ Multilayer System: Rp and Rs versus incident angle for both effective medium theory (dashed) and the multilayer structure (solid) for Type I ($\lambda = 360$ nm) and Type II ($\lambda = 750$ nm) hyperbolic regions. The structures are at 50% metal fill fraction for the both the EMT slab (320 nm thick) and the 40 layered multilayer structure (8 nm layers). (b) Rp and Rs versus incident angle for Type I ($\lambda = 850$ nm) and Type II ($\lambda = 390$ nm) Ag-Al$_2$O$_3$ nanowire system (500 nm thick) are also shown for 15 % metal fill fraction. Embedded schematic in (b) shows the defined incident angle ($\theta$).

The Brewster angle, the angle for which there is a strong minimum in the reflectance, is seen clearly for the p-polarized Type I reflectance in both the multilayer (figure 5(a)) and nanowire (figure 5(b)) structures. Free electron motion parallel to the interface greatly increases the overall reflection for Type II hyperbolic metamaterials and a Brewster angle cannot be defined. The behavior of the Brewster angle with respect to the wavelength can be used to determine whether the metamaterial exhibits Type I behavior by looking at the propagating spectrum. A discontinuity in the Brewster angle is witnessed at the wavelength where $\varepsilon_\perp \approx 0$ in a Type I hyperbolic metamaterial [24].

*Transmission spectrum*: As discussed, the Type I regime shows high transmission in contrast to other metamaterials where absorption is a major issue. In figure 6(a) we plot the transmission from a multilayer sample which shows a window of transparency in the Type I region until it becomes very reflective in the Type II regime. The effective medium theory predictions are well matched to the multilayer simulations using an Ag-TiO$_2$ system. Figure 6(b) shows the transmission through the nanowire Type I region. Moving the transmission windows of hyperbolic media to visible wavelengths is a challenge with the multilayer design unless high index dielectrics are used.

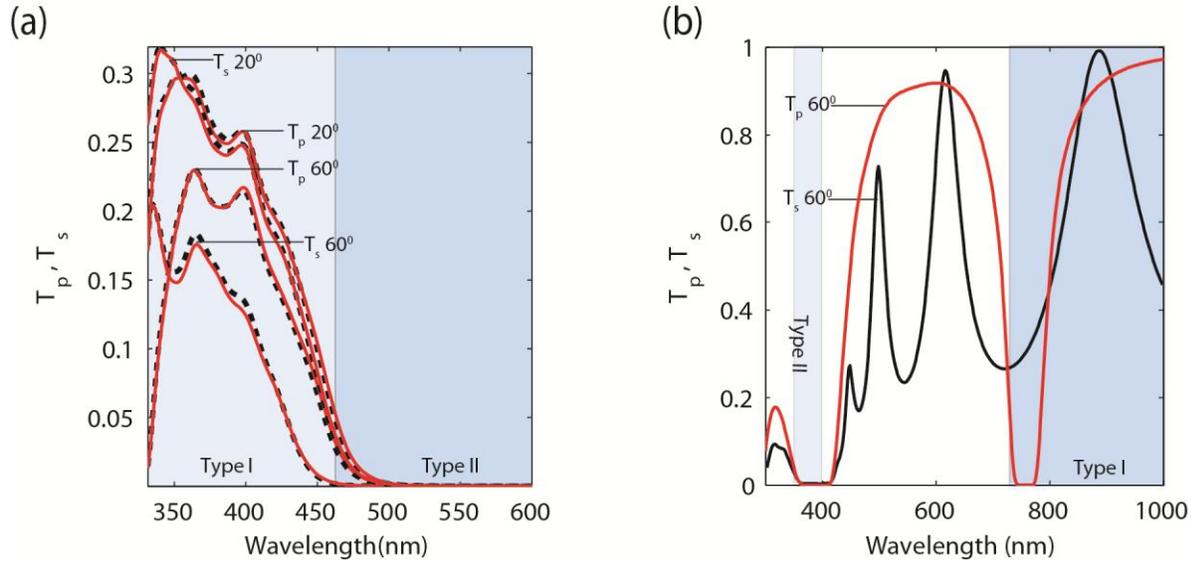

Figure 6: (a) Multilayer System: Transmission versus wavelength for an Ag-TiO$_2$ system for both p and s polarized light at different incidence angles with effective medium theory (dashed) and a multilayer structure (solid). The structures are at 50% metal fill fraction for the both the EMT slab (320 nm thick) and the 40 layered multilayer structure (8 nm layers). (b) Nanowire System: Transmission versus wavelength for p and s polarizations at a 60$^0$ incident angle for an Ag-Al$_2$0$_3$ nanowire slab (500 nm thick) at 10 % fill fraction. The wavelength range shown spans the Type I and Type II hyperbolic regions in both (a) and (b).

*Extinction spectrum:* The nanowire design shows interesting features in the extinction spectrum which arises due to the epsilon-near-zero (ENZ) and epsilon-near-pole (ENP) resonances (figure 7). The ENZ resonance requires a component perpendicular to the interface and occurs only for p-polarized light. In the nanowire structure, the directions of ENZ and ENP resonances are such that they interact with propagating waves and subsequently lead to large extinction. Specifically, from the displacement boundary condition, $\varepsilon_0 E_{0\perp} = \varepsilon_\perp E_{1\perp}$, and therefore when $\varepsilon_\perp \to 0$, the fields inside the nanowire HMM ( $E_{1\perp}$ ) should be very large. Thus large absorption is expected at this epsilon near zero region for this particular nanowire structure. Although this resonance is seen in the multilayer structure, it has free electron motion parallel to the interface and thus greatly reflects the incoming propagating fields. The ENZ and ENP resonances, therefore, do not appear in the multilayer extinction spectrum due to the characteristically higher reflection from increased metallic behavior of such structures.

The ENP resonance occurs for both polarizations. They are often addressed in literature as the L (longitudinal) and T (transverse) resonance corresponding to the direction of plasmonic oscillations in the rod [58].

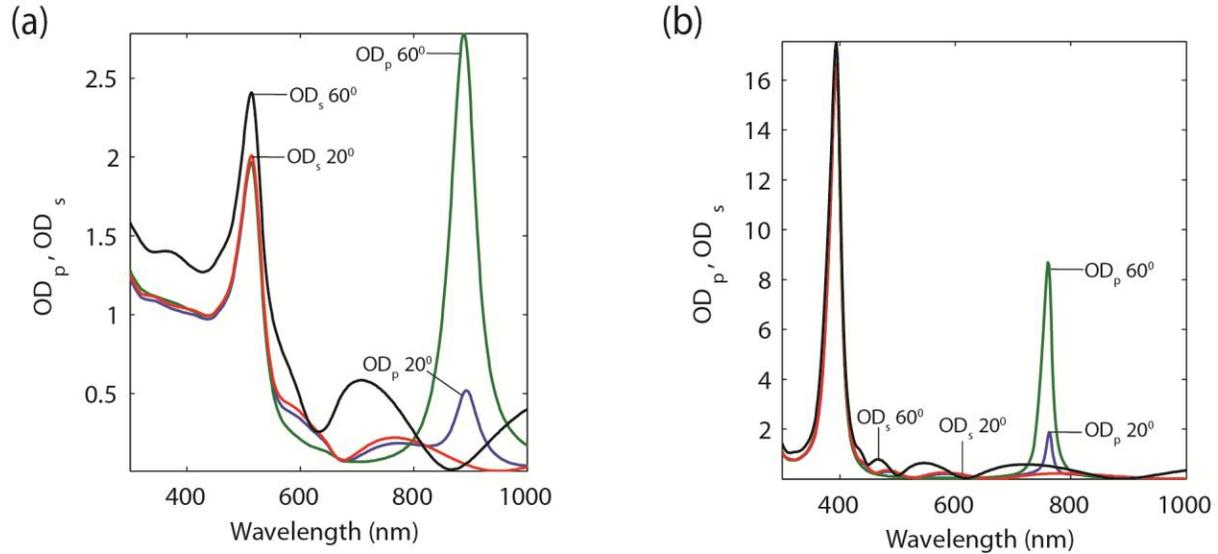

Figure 7: Optical Density (OD) versus wavelength for both p and s incident polarizations for an Au-Al$_2$O$_3$ system (a) and an Ag-Al$_2$O$_3$ system (b) both at 10% metal fill fraction (500 nm slab thickness). The T resonance is shown in (a) at $\lambda = 500$ nm and the L resonance at $\lambda = 900$ nm. In (b), the T resonance occurs at $\lambda = 400$ nm and the L resonance at $\lambda = 775$ nm. Only p-polarized incidence shows the L resonance. Note that the resonances in silver are stronger due to the relatively lower losses. These extinction resonances only occur in the nanowire geometry and not the multilayer structures.

c) Evanescent wave spectrum

The most important characteristic of hyperbolic behavior cannot be discerned by propagating waves alone. This is because multiple applications stem from the high-$k$ propagating waves in the medium which are evanescent in vacuum. An optical tunneling experiment is essential to understand the high-$k$ waves [64]–[66]. However a significant issue is that even such an experiment would require high index prisms for in and out-coupling which are not readily available in the visible range. A grating based approach is ideal to study these high-$k$ waves [37].

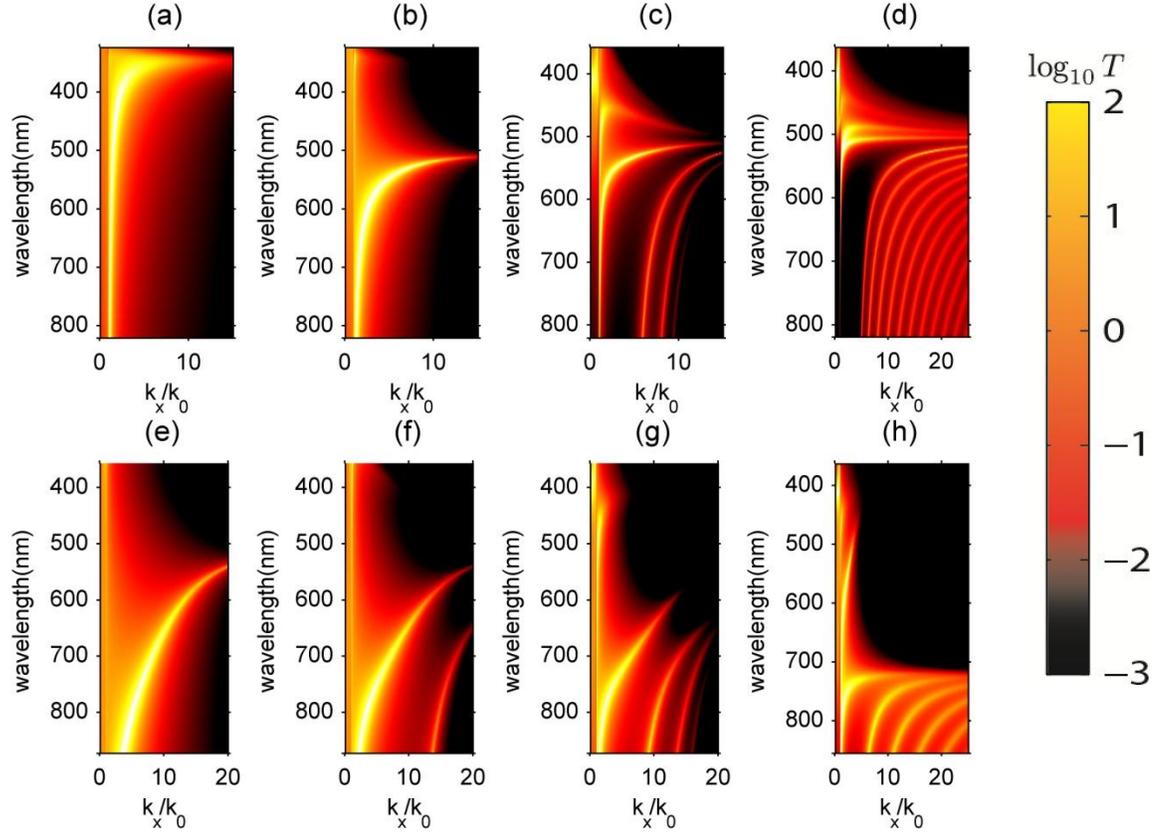

Figure 8: Calculated transmission (log scale) through various Ag-TiO$_2$ systems using the transfer matrix method. (a) 1 30 nm Ag layer. Ag (30 nm) and TiO$_2$ (30 nm) with 2 alternating layers (b), 8 alternating layers (c) and a 240 nm thick EMT slab with 50 % fill fraction (d). Ag (10 nm) and TiO$_2$ (30 nm) for 2 alternating layers (e), 4 alternating layers (f), 8 alternating layers (g) and a 240 nm thick EMT slab 25 % fill fraction (h).

Here we study the transmission of evanescent waves incident on the multilayer realization of the HMM using the transfer matrix method. Similar studies have also looked at alternative approaches to view evanescent wave spectra [67]. The evanescent waves are labeled by their wavevector parallel to the vacuum-HMM interface ($k_x > k_0 = \omega / c$). The propagating waves have $k_x < k_0$. Figure 8 shows the effect of increasing the number of layers of silver in the multilayer structure on the transmission spectrum across the visible range. For the thin layer of silver in vacuum we see the bright band corresponding to the characteristic surface plasmon polariton dispersion. When a thin layer of dielectric is added, this bilayer system has an interface plasmon polariton with a shifted resonance frequency as shown in figure 8(b). Figure 8(c) shows the scenario when there are 8 alternating layers of silver and $TiO_2$. Interestingly, multiple bands of interface waves are seen. This has to be compared to figure 8(d) where the effective medium simulation has been shown for a slab of the same total thickness. It thus becomes evident that the high-*k* waves, which can be interpreted as high-*k* waveguide modes in the slab geometry, originate from the plasmon polaritons at the interface of silver and $TiO_2$. The coupling between multiple such polaritons leads to a splitting of modes to higher wavevectors and energies. This is evident in the emergence of new modes as the number of layers are increased. In figure 8(e)-(h) we show the case of polaritonic high-*k* waves with a $TiO_2$ thickness of 30 nm and a 10 nm silver thickness for various numbers of alternating layers. It is clear that the increased dielectric ratio (25 % fill fraction) of the structure changes the nature of the dispersion of the high-*k* modes and causes a subsequent shift in the plasma frequency. We see no high-*k* waves in the elliptical dispersion range for this subwavelength unit cell structure [57]. However, it should be noted that some studies have shown high-*k* multilayer plasmons in the elliptical dispersion regime [68].

**5.0 Applications**

a) Far-field sub-wavelength imaging

One of the major applications of metamaterials has been in the area of subwavelength imaging. Conventional optics is known to be limited by the diffraction limit, i.e., the ability of a conventional lens to focus light or form images is always constrained by the wavelength of the illuminating light.

We first revisit the conventional diffraction limit by understanding the behavior of light scattering from an object. For the sake of discussion we limit ourselves to the 2D case. When light scatters off an object, the far-field light does not capture its sharp spatial features. The image of the object constructed from the far-field light loses these parts of the image as shown in figure 9, which in other words can be interpreted as the diffraction limit [1].

Note these large wavevector waves carry spatial information about the subwavelength features of the object and decay exponentially. This is due to the spatial bandwidth of vacuum which allows propagating waves with wavevector $k_x < k_0$ (figure 9(b)).

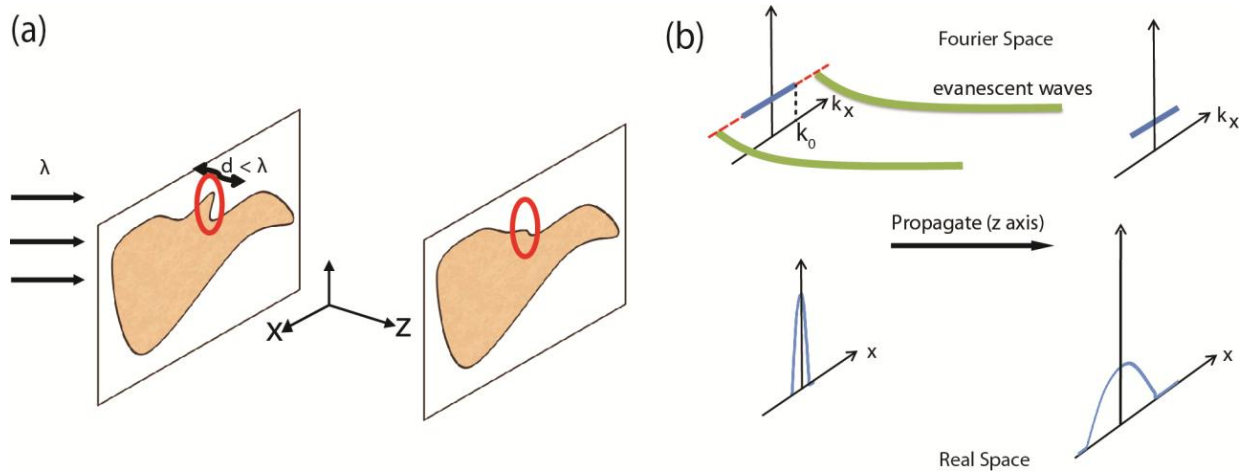

Figure 9: (a) Far-field imaging: Light scattered off an object loses sharper spatial features in the far-field due to the diffraction limit. (b) Larger wavevectors carry the subwavelength spatial features of the object ($k_x > k_0$). These large wavevectors decay exponentially (evanescent waves) in vacuum due to the limited spatial bandwidth and, as a result, finer features of the image are lost. The loss of information in both fourier space and real space are shown.

However, when a multilayer hyperbolic metamaterial is brought to the near-field of the object, these waves which are evanescent can propagate in the medium and the image is translated with subwavelength resolution to the output face of the metamaterial. This can be understood by observing the behavior of point dipole radiation near a hyperbolic medium as shown in the inset of figure 10(a). Instead of the conventional dipole radiation pattern, it is seen that the dipole radiates into sub-diffraction resonance cones [15], [18], [69]. Each point on the object can be considered as a radiator and the pixels of information are translated by the resonance cones with sub-diffraction resolution to the output interface. However, a major drawback is that the wave is evanescent outside the metamaterial and cannot carry information to the far-field [70].

The hyperlens is a device which overcomes this limitation [3], [25], [71], [72]. Evanescent wave energy and information from the near-field can be transferred to the far-field if the layers forming the hyperbolic metamaterial are curved in a cylindrical fashion. The qualitative explanation for this conversion phenomenon is shown in figure 10(b). Conservation of angular momentum ($m \sim k_\theta r$) in the cylindrical geometry implies that the tangential part of the momentum for the wave times the radius is a constant. Thus when the high-$k$ waves move towards the outer edge of the cylinder, the wavevectors decrease in magnitude. If the radii are carefully chosen, then the wavevectors can be compressed enough to propagate to the far-field. This enables far-field subwavelength resolution using the hyperlens [72], [73].

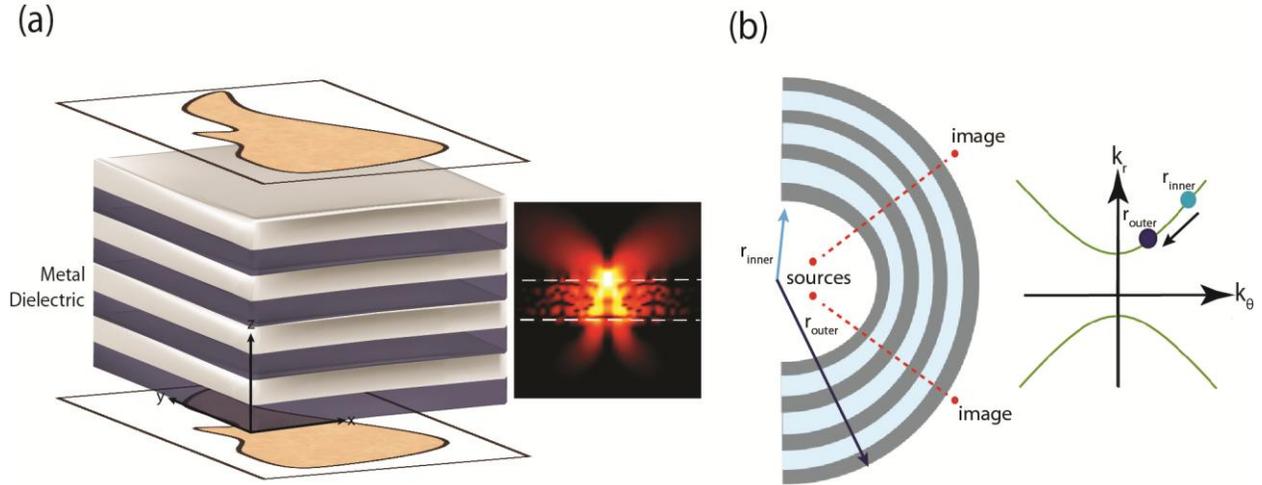

Figure 10: Subwavelength Imaging with HMMs: (a) Multilayer HMM in the near field of an object allows normally evanescent waves to propagate and carry subwavelength features across the length of the structure. Inset: Point dipole placed in the near field of a Type I HMM. The dipole radiates into sub-diffraction resonance cones in the HMM structure. The high-*k* waves are still evanescent outside of the multilayer HMM. (b) Hyperlens: Cylindrical HMM geometry allows for the tangential component of the wavevector times the radius to remain constant. The wavevectors decrease in magnitude as the high-*k* waves move to the edge of the structure. If the wavevector magnitudes are reduced sufficiently to allow propagation in vacuum, the hyperlens can carry subwavelength features to the far-field.

An important consideration for practical realization and optimization is the additional impedance matching condition imposed on the metal and dielectric building blocks [40], [74]. This condition gives ( $\mathrm{Re}(\varepsilon_m + \varepsilon_d) \approx 1$ ) for equal layer thicknesses [75]. Note the hyperlens functions in the Type I HMM regime since it requires high transmission. Currently, there are major efforts underway to make the hyperlens into the first metamaterial device with practical applicability in imaging systems.

b) Density of states engineering

An orthogonal direction of application for hyperbolic media is in the area of engineering the photonic density of states (PDOS) [21], [22], [27], [76], [77]. A critical effect was unraveled with regards to the density of electromagnetic states inside hyperbolic media using the analogy with the electronic density of states (EDOS). The EDOS is calculated by computing the volume enclosed between Fermi surfaces with slightly different energies. For closed surfaces such as spheres and ellipsoids this calculation leads to a finite value. In the case of the PDOS of the hyperbolic medium, it is clearly seen that this volume diverges leading to an infinite density of electromagnetic states within the medium (figure 11(a)).

Fermi's golden rule states that the spontaneous emission lifetime of emitters is strongly influenced by the density of available electromagnetic modes [78]. When fluorescent dye molecules or quantum dots are brought near the hyperbolic metamaterial the interaction is dominated by the modes with the highest density of states. As compared to the modes in vacuum, the hyperbolic high-$k$ states dominate and the emitters preferentially couple to these modes [57]. This leads to a decrease in lifetime. Multiple experiments have explored this effect by studying dye molecules and quantum dots on top of the multilayer and nanowire hyperbolic metamaterial. A lifetime shortening has been observed but discerning the radiative and non-radiative effects have proven to be a challenge [19]. Here, we outline the factors underlying the multitude of experiments measuring spontaneous emission lifetime.

   i.   *Absorption enhancement:* One cause of fluorescence enhancement is the increase in the absorption of molecules or quantum dots due to the effect of the environment. The absorption increases due to local field enhancement and is proportional to the local field intensity at the location of the emitter ( $A \propto |E|^2$ ). It is necessary to keep the absorption constant across samples to reliably compare the photoluminescence enhancement [79]. One approach to achieving this is to function in the saturation regime, where the emitted power is no longer proportional to the input power. In this illumination regime, the fluorophores absorb the maximum possible power leading to a constant steady state excited population for pulsed excitation. Any increase in photoluminescence in this regime can be attributed to enhanced decay rates due to the environment and not absorption enhancement [80], [81].

For the case of planar multilayer hyperbolic metamaterials, the absorption enhancement is expected to be a weak effect due to the lack of any localized plasmons which generate field hotspots [82]. Furthermore, the high-*k* modes cannot be excited by free space illumination and cannot affect the absorption.

ii.  *Photoluminescence enhancement:* In the regime where the output fluorescence signal is independent of the input power (saturation regime), the photoluminescence can be enhanced by increasing the radiative decay rate [83]. This rate depends on the available photonic density of states. Increasing the radiative decay routes for the emitter through the enhancement of the PDOS can be accomplished using microcavities, photonic crystals, plasmonics or metamaterials.

The increase in the radiative decay rate is known as the Purcell factor defined as

$F_P = \Gamma_{rad}^{env} / \Gamma_{rad}^0$ where $\Gamma_{rad}^{env}$ is the enhanced radiative decay rate due to the photonic environment and $\Gamma_{rad}^0$ is the radiative decay rate in vacuum. Notice the environment might also introduce non-radiative channels of decay, and in vacuum the emitter might have intrinsic non-radiative decay channels. Hence, the definition utilizes only radiative decay rates as the correct measure of photoluminescence enhancement. It should be noted that the quantities defined are related to the signal intensity measured at the detector. We can only measure the net decay rate and infer the radiative rate from the PL intensity data with and without the environment [84].

The Purcell factor can also be defined for a mode through the density of states $F_p = \rho^{env}(\omega_0) / \rho^0(\omega_0)$. This theoretical definition does not directly imply a large PL enhancement since the out-coupling factor for the mode into vacuum might be low. For example, a surface plasmon polariton might have a large density of states but unless this surface mode is out-coupled to the detector, the measured PL enhancement will be minimal [57], [85]. It is important to note, therefore, that if one wants to determine the exact PL enhancement, an out-coupling efficiency should be associated with the mode that potentially allows the increased enhancement [86]. It is necessary while comparing theory and experiment to carefully isolate the out-coupling factor and also take into account the overlap of the emitter to the field profile of the mode.

In the case of hyperbolic metamaterials, defining a unique modal Purcell factor through the density of states is not possible since the high-$k$ modes form a continuum of available states. Nevertheless, one can define a net enhancement in the density of states as well as PL enhancement [86].

iii.  *Radiative efficiency:* For applications such as single photon sources, the Purcell factor, photoluminescence enhancement, and collection efficiency form the key figures of merit [86]–[88]. However, for light emitting diodes the radiative efficiency of the source is essential (i.e. the output photoluminescence for a given input power [83]). A large Purcell factor always implies photoluminescence enhancement (assuming efficient out-coupling of light) since the excited state can relax faster radiatively. The saturation pump power also increases due to the Purcell factor. However, PL enhancement does not necessarily imply radiative efficiency enhancement since it often comes at the cost of larger input powers.

To understand this, we define the intrinsic ($\eta_i$) and apparent quantum yield ($\eta_a$) of the emitter as

$$\eta_i = \frac{\Gamma_{rad}^0}{\Gamma_{rad}^0 + \Gamma_{non-rad}^0} \quad (2)$$

and

$$\eta_a = \frac{\Gamma_{rad}^{env}}{\Gamma_{rad}^{env} + \Gamma_{non-rad}^{env} + \Gamma_{non-rad}^0} \quad (3)$$

The intrinsic quantum yield measures the internal radiative efficiency of the emitter and the apparent quantum yield is the final radiative efficiency in the presence of the nanostructures. Note we assume the internal radiative efficiency is constant for interaction distances further than 5 nm.

The radiative efficiency enhancement is defined as:

$$F = \frac{\eta_a}{\eta_i} = \frac{\Gamma_{rad}^{env}(\Gamma_{rad}^0 + \Gamma_{non-rad}^0)}{\Gamma_{rad}^0(\Gamma_{rad}^{env} + \Gamma_{non-rad}^{env} + \Gamma_{non-rad}^0)} \quad (4)$$

Here, F is the ratio of the apparent to intrinsic quantum yields. The plasmonic or metamaterial environment invariably increases both the radiative and non-radiative rates. Therefore, the radiative efficiency enhancement can only be substantial for emitters with a low intrinsic quantum yield. This is detrimental to applications in LEDs where the primary figure of merit is the radiation efficiency increase.

iv. *Decay rate increase/Lifetime shortening*: Time resolved measurements of the spontaneous emission give access to the net decay rate or lifetime of the fluorophore. This rate is always the sum of multiple factors and in particular, the total lifetime ($\tau$) near the hyperbolic metamaterial can be written as

$$\frac{1}{\tau} = \frac{1}{\tau_{rad}} + \frac{1}{\tau_{non-rad}} \quad (5)$$

where $\tau_{rad}$ is the radiative lifetime and $\tau_{non-rad}$ is the non-radiative lifetime. Note that the decrease in total lifetime is not indicative of photoluminescence enhancement since the modes into which the light couples are dark (i.e. they don't out-couple to vacuum). Natural surface roughness of the layers or gratings are needed to out-couple this light to the detector in order to calculate the radiative lifetime experimentally. Figure 11(b) shows the decrease in lifetime of an emitter as it is moved closer to a metamaterial surface [19], [28], [89], [90].

v.   *Quenching:* We now discuss the concept of quenching near metallic structures and hyperbolic metamaterials in particular [91]. In the dipolar approximation, the fluorophores are treated as point emitters with waves of all spatial harmonics emanating from it at the transition frequency. In the near-field of any absorptive structure, the waves with large spatial harmonics are simply absorbed [92], [93]. Since they are the dominant route as compared to vacuum modes for near-field interaction, lifetime decrease occurs primarily due to non-radiative decay. This phenomenon of reduced photoluminescence is known as quenching.

We emphasize that quenching itself is a phenomenon that depends on the local density of states and once competing channels are available at the large wavevectors the non-radiative decay can be overcome. This phenomenon is depicted when comparing figure 11(c) and figure 11(d) where we show the power spectrum of light emitted by fluorophore at a dye distance of 3 nm and 20 nm, respectively. The power spectrum is analogous to the wavevector resolved local density of states. It is seen that when the dye is very close to the surface of the HMM or the silver slab (figure 11(d)) we see a smooth peak at larger magnitudes of the in plane wavevector ($k_x$). This smooth peak is a result of quenching and does not correspond to propagating modes with a well-defined dispersion. One notes that this smooth peak is not evident when the dipole is placed farther away from the structures (figure 11(c)) where the effects of quenching are reduced. Note that in the effective medium limit, for both the Type I and the Type II structures, we still see the peaks in the LDOS corresponding to the high-$k$ modes of the structure when quenching is present. Thus, the nature of the unbounded wavevectors for the hyperbolic metamaterial allow for propagating modes to exist even in the regime where larger quenching effects are taking place. It is interesting to note that silver seems to show a larger decrease to the emitter lifetime when the emitter is placed extremely close to the surface (figure 11 (b)). This can be attributed to a larger smooth quenching peak than the Type I HMM (figure 11(d)). Type II metamaterials show a larger number of propagating high-$k$ modes.

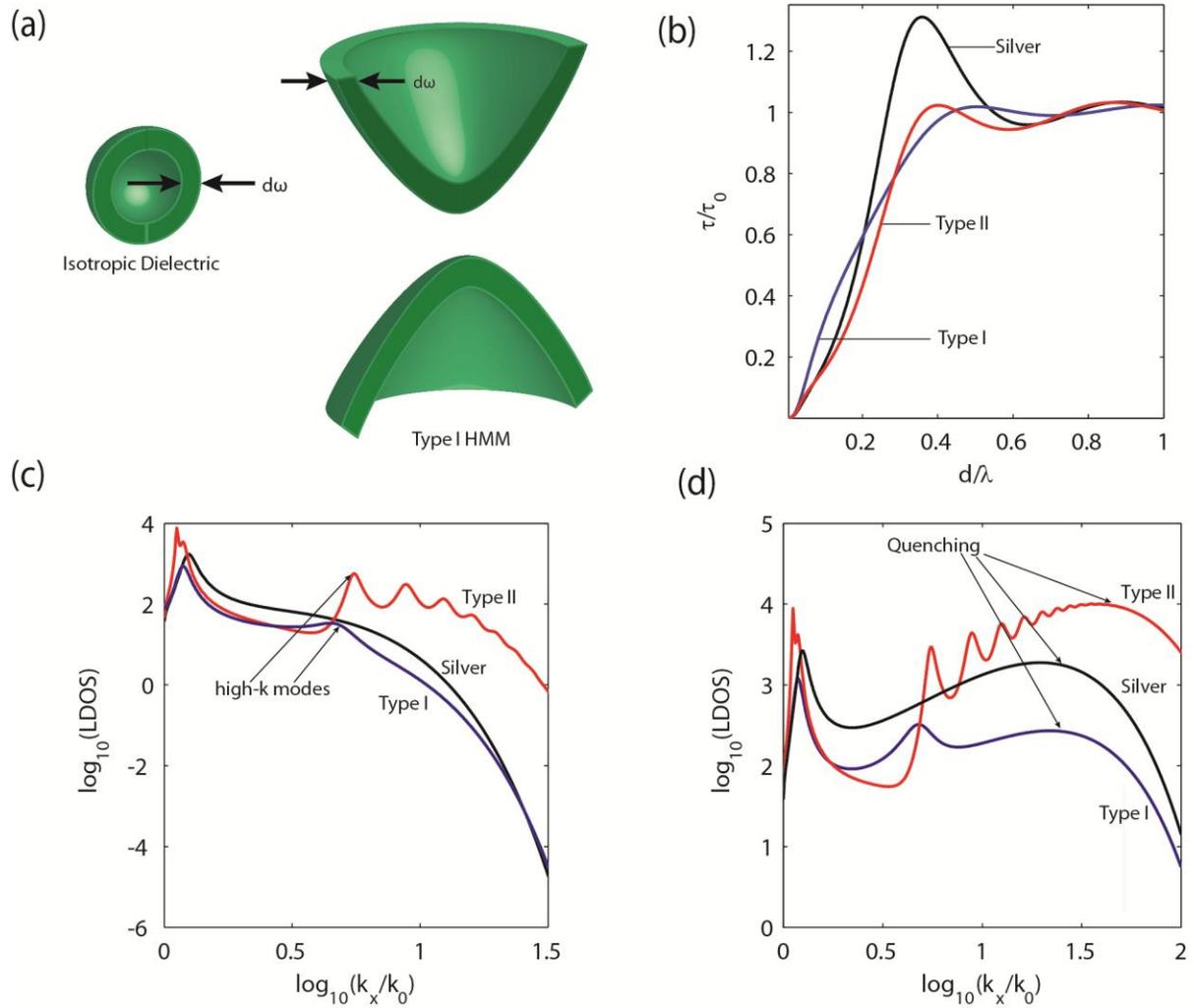

Figure 11: (a) Isofrequency surfaces at slightly different energies for an isotropic dielectric and a Type I HMM. The enclosed volume between the two isofrequency surfaces is a measure of the photonic density of states of the system. It is clear that the HMM has a diverging enclosed volume and thus, in the ideal limit, can support an infinite photonic density of states. (b) Lifetime of a dipole, normalized to the free space lifetime, versus distance above the HMM surface ("*d*"). An Ag-TiO$_2$ system with 35% fill fraction is considered in the Type I ($\lambda = 350$ nm) and Type II ($\lambda = 645$ nm) regions. A thick film of silver ($\lambda = 372$ nm) is also shown for comparison. Local density of states (LDOS) versus wavevector for a 200 nm Ag-TiO$_2$ slab (35 % fill fraction) and 200 nm silver film for an emitter placed (c) 20 nm and (d) 3 nm above the structure. Note that high-*k* modes exist in both (c) and (d) however a clear broad quenching peak is seen in (d).

## 6.0 Conclusion and Future Work

In conclusion, we have reviewed the major applications of hyperbolic media: subwavelength imaging and photonic density of states engineering. To aid experimentalists, we have described the practical approaches of designing and characterizing thin film and nanowire hyperbolic metamaterials. The major future work in this area will be about engineering the coherent, thermal and quantum state of light. These media present a unified platform for building nanoscale light emitters from nanoscale lasers [94], [95] to broadband super-Planckian thermal emitters [32]. Another major direction will be the fluctuational and macroscopic quantum electrodynamics of metamaterials [96]. Hyperbolic media have become an important class of artificial photonic materials for research and is expected to be the first optical metamaterial to find widespread applicability in device applications.

# Appendix 1.0: Effective Medium Theory for a Multilayer System

Here, we will look at deriving the effective medium permittivities for an anisotropic multilayer composite with a uniaxial symmetry. The method follows a generalized Maxwell-Garnett approach to obtain analytical expressions for the effective permittivity in the parallel ($\varepsilon_\parallel$) and perpendicular ($\varepsilon_\perp$) directions defined below for the multilayer metamaterial.

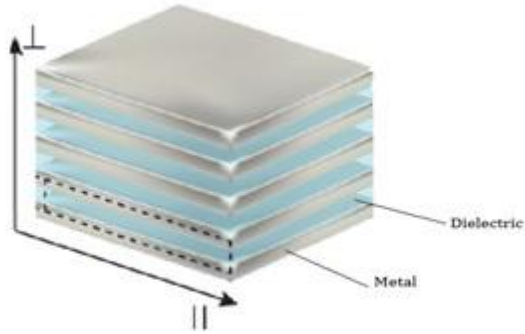

Figure 1: Schematic of anisotropic multilayer composite. The perpendicular direction is defined as parallel to the normal vector from the surface of the metamaterial and the parallel direction is defined as the plane parallel to the metamaterial interface.

Figure 1 outlines the nature of the alternating metallic and dielectric layers to form a multilayer structure. The metallic and dielectric layers have permittivities $\varepsilon_m$ and $\varepsilon_d$ respectively. Furthermore, we can define the fill fraction of the total thickness of metal in the system to the total thickness of the metamaterial as follows:

$$\rho = \frac{d_m}{d_m + d_d} \quad (1)$$

where $d_m$ is the sum of all the thicknesses of metallic layers in the system and $d_d$ is the sum of all the thicknesses of the dielectric layers.

## Effective Parallel Permittivity:

In this section, we will derive our analytical expression for the parallel component of the permittivity tensor of our multilayer system. We can start by noting that the electric field displacement ($D$) is proportional to the electric field ($E$) through the following equation:

$$\vec{D} = \overline{\overline{\varepsilon}}_{eff} \vec{E} \quad (2)$$

where $\varepsilon_{eff}$ is the overall effective permittivity of the medium. We know from electrostatics that the tangential component of the electric field must be continuous across an interface as we go from one medium to another. Therefore, we can say

$$E_m^{\parallel} = E_d^{\parallel} = E^{\parallel} \quad (3)$$

where we can take $E_m^{\parallel}$ to be the electric field in the metallic layers, $E_d^{\parallel}$ to be the electric field in the dielectric layers, and $E^{\parallel}$ is the electric field of the subwavelength metamaterial. From the continuity of the dielectric displacement in the parallel direction explained above, we can find the overall displacement by averaging the displacement field contributions from the metallic and dielectric components:

$$D^{\parallel} = \rho D_m^{\parallel} + (1-\rho) D_d^{\parallel} \quad (4)$$

Substituting equations (2) and (3) to the above, we get:

$$\varepsilon_{eff}^{\parallel} E^{\parallel} = \rho \varepsilon_m E^{\parallel} + (1-\rho) \varepsilon_d E^{\parallel} \quad (5)$$

If we cancel out the common parallel electric field components, we arrive at the final equation:

$$\varepsilon_{\parallel} = \rho\varepsilon_m + (1-\rho)\varepsilon_d \quad (6)$$

## Effective Perpendicular Permittivity:

To derive our expression for the perpendicular permittivity, we can again start from Maxwell's Equations and use electromagnetic field boundary conditions. We specifically know that the normal component of the electric displacement vector at an interface must be continuous which gives us the expression

$$D_m^{\perp} = D_d^{\perp} = D^{\perp} \quad (7)$$

We also know that the total magnitude of the electric field will be a superposition of the electric field components from the dielectric layers and the metallic layers. Thus, we can define

$$E^{\perp} = \rho E_m^{\perp} + (1-\rho)E_d^{\perp} \quad (8)$$

where $E_m^{\perp}$ is the perpendicular component of the electric field in the metallic region, $E_d^{\perp}$ is the perpendicular component of the electric field in the dielectric region, and $E^{\perp}$ is the total electric field in the multilayer system. We can now use our boundary condition from equation (7) and Maxwell's equation from (2) and substitute them into equation 8. If we cancel out the common electric field terms and solve for $\varepsilon_{\perp}$, we find the analytic expression for the electric permittivity of the multilayer metamaterial in the perpendicular direction:

$$\varepsilon_{\perp} = \frac{\varepsilon_m\varepsilon_d}{\rho\varepsilon_d + (1-\rho)\varepsilon_m} \quad (9)$$

## Appendix 2.0: Effective Medium Theory for a Nanowire System

Here we will look at deriving the effective medium permittivities for an anisotropic nanowire composite with a uniaxial symmetry. The method follows a generalized Maxell-Garnett approach to obtain analytical expressions for the effective permittivity in the parallel ($\varepsilon_\parallel$) and perpendicular ($\varepsilon_\perp$) directions of the nanowire metamaterial.

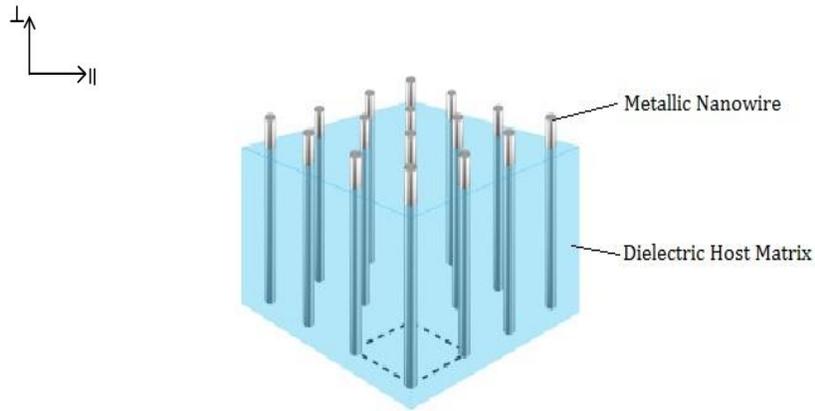

Figure 1: Schematic of an anisotropic nanowire composite. The perpendicular direction is defined along the long axis of the nanowire and the parallel direction is defined as the plane along the short axis of the nanowire.

Figure 1 outlines the nature of the embedded metallic nanowires in the dielectric host which have defined permittivities $\epsilon_m$ and $\epsilon_d$ respectively. Furthermore, we can define the fill fraction of nanowires ($\rho$) in the host material as follows:

$$\rho = \frac{\text{nanowire area}}{\text{unit cell area}} = \frac{a}{A} \tag{1}$$

The hexagonal unit cell geometry consists of 3 nanowires per unit cell (1 centre wire plus additional partial nanowires at each vertex of the hexagon). $A$ is the unit cell area of the hexagon and $a$ is the cross sectional area of a single metallic nanowire.

## Effective Parallel Permittivity

In this section we will derive our analytical expression for the parallel component of the permittivity tensor for our nanowire system. We can start from Schrodinger's Wave Equation, which has a solution as a function of the radial distance from the centre of the nanowire ($r$). We note that at a distance $R$ (the radius of the nanowire) we approach the metallic nanowire and dielectric host boundary. Defining our potential inside the nanowire as $\psi_1$ and the potential of the dielectric host as $\psi_2$, we can make the following assertions about the limits of our potential as well as their behavior at the boundary from known boundary conditions:

$$\psi_1 \mid_{r=R} = \psi_2 \mid_{r=R} \tag{2}$$

$$\mid \psi_1(r=0) \mid < +\infty \tag{3}$$

$$\mid \psi_2(r \to \infty) \mid = -E_o r \cos\theta \tag{4}$$

$$\varepsilon_1 \frac{d\psi_1}{dr} \mid_{r=R} = \varepsilon_2 \frac{d\psi_2}{dr} \mid_{r=R} \tag{5}$$

In equation 5, $\varepsilon_1$ and $\varepsilon_2$ are the permittivities inside the metallic nanowire and the dielectric host respectively. We can suggest an arbitrary solution for $\psi$ using a trigonometric series expansion for the periodic potential:

$$\psi = A\ln r + K + \sum_{n=1}^{\infty} r^n (A_n \sin n\theta + B_n \cos n\theta) + \sum_{n=1}^{\infty} \frac{1}{r^n}(C_n \sin n\theta + D_n \cos n\theta) \tag{6}$$

Now, using our conditions outlined in equations 2-5, we can make approximations to define the potential functions $\psi_1$ (inside the nanowire) and $\psi_2$ (outside the nanowire) as the following:

$$\psi_1 = K_1 + \sum_{n=1}^{\infty} r^n (A_n \sin n\theta + B_n \cos n\theta) \qquad (7)$$

$$\psi_2 = K_2 - E_o r \cos\theta + \sum_{n=1}^{\infty} \frac{1}{r^n} (C_n \sin n\theta + D_n \cos n\theta) \qquad (8)$$

We drop the $\frac{1}{r^n}$ term in $\psi_1$ (equation 7) and replace the $r^n$ term in $\psi_2$ (equation 8) with the limit of the potential at $\psi_2 \to \infty = -E_o r \cos\theta$. This ensures non-infinite solutions for the potentials at all values of *r*. Furthermore, we can set the values of the constants $K_1$ and $K_2$ in equation 7 and equation 8 to 0. This is due to the fact that when we take the derivative of the potential ($\psi$) to eventually find our electric fields, these constants will subsequently disappear.

We can now use our boundary condition given in equation 2 for the interface between the nanowire and the dielectric host at R with our defined potentials inside the nanowire ($\psi_1$) and outside the nanowire ($\psi_2$). Due to the uniqueness of this trigonometric series expansion we can equate the coefficients of the trigonometric functions for our expression at the boundary.

$$A_n R^n = \frac{C_n}{R^n} \qquad (9)$$

$$B_n R^n = \frac{D_n}{R^n} \qquad (10)$$

$$R B_1 = \frac{D_1}{R} - E_o R \qquad (11)$$

Equation 11 is the relation between the coefficients when n =1 in our expansion. We can also

write an expression for our second boundary condition given by plugging in equation 7 and

equation 8 into equation 5:

$$\varepsilon_1 \sum_{n=1}^{\infty} A_n n R^{n-1} (A_n \sin n\theta + B_n \cos n\theta) = -\varepsilon_2 E_o \cos \theta - \varepsilon_2 \sum_{n=1}^{\infty} \frac{n}{R^{n+1}} (C_n \sin n\theta + D_n \cos n\theta) \quad (12)$$

We can once again equate the coefficients of the trigonometric functions given by equation 12 to obtain 3 new relations:

$$\varepsilon_1 n A_n R^{n-1} = \varepsilon_2 \frac{-n}{R^{n+1}} C_n \quad (13)$$

$$\varepsilon_1 n B_n R^{n-1} = \varepsilon_2 \frac{-n}{R^{n+1}} D_n \quad (14)$$

$$\varepsilon_1 B_1 = -\varepsilon_2 E_o - \varepsilon_2 \frac{D_1}{R^2} \quad (15)$$

We now note that we can set $A_n = B_n = C_n = D_n = 0$ because they give impossible boundary conditions. However, we can still use equation 11 and equation 15 to solve for the coefficients $D_1$ and $B_1$ through substitution:

$$D_1 = \frac{\varepsilon_1 - \varepsilon_2}{\varepsilon_1 + \varepsilon_2} R^2 E_o \quad (16)$$

$$B_1 = \frac{-2\varepsilon_2}{\varepsilon_1 + \varepsilon_2} E_o \quad (17)$$

Equation 17 now gives us our expression for $B_1$ which we can substitute into our expression for $\psi_1$ (equation 7) at n = 1:

$$\psi_1 = \frac{-2\varepsilon_2}{\varepsilon_1 + \varepsilon_2} E_o R \cos \theta \quad (18)$$

Equation 18 now gives us our expression for the potential inside the well in terms of the electric field outside of the nanowire ($E_0 = E_{out}$). We can differentiate equation 18 with respect to R to get our expression for the electric field inside the nanowire ($E_{in}$).

$$-\frac{\partial(\psi_1)}{\partial R} = E_{in} = \frac{2\varepsilon_2}{\varepsilon_1 + \varepsilon_2} E_{out} \tag{19}$$

The known isotropic relation for the parallel permittivity ($\varepsilon_{\parallel}$) with two different material mediums is given by:

$$\varepsilon_{\parallel} = \frac{\rho \varepsilon_1 E_{in} + (1-\rho)\varepsilon_2 E_{out}}{\rho E_{in} + (1-\rho) E_{out}} \tag{20}$$

We can now substitute our expression for $E_{in}$ (equation 19) into equation 20 and make the further substitution that $\varepsilon_1 = \varepsilon_m$ and $\varepsilon_2 = \varepsilon_d$ corresponding to the permittivity of the metallic nanorod and dielectric host respectively. Upon making this substitution we arrive at our expression for the parallel component of the permittivity tensor ($\varepsilon_{\parallel}$) in terms of the metallic nanorod fill fraction ($\rho$) and the permittivities of the nanorod ($\varepsilon_m$) and dielectric host ($\varepsilon_d$)

$$\varepsilon_{\parallel} = \frac{(1+\rho)\varepsilon_m \varepsilon_d + (1-\rho)\varepsilon_d^2}{(1+\rho)\varepsilon_d + (1-\rho)\varepsilon_m} \tag{21}$$

## Effective Perpendicular Permittivity

We can derive our expression for the perpendicular permittivity from Maxwell's Equations and making use of the electromagnetic boundary conditions. Specifically, we know that the tangential component of the electric field ($E^{\perp}$) along the long axis of the nanowire is continuous at the boundary between the nanowire and the dielectric host.

$$E_1^\perp = E_2^\perp = E^\perp \tag{22}$$

Here $E_1^\perp$ is the perpendicular electric field in the metallic nanowire, $E_2^\perp$ is the perpendicular electric field in the dielectric host and $E^\perp$ is the effective perpendicular field for the subwavelength nanowire metamaterial. We note from Maxwell's Equations that the displacement field in the perpendicular direction can be defined as $D^\perp = \varepsilon_\perp E^\perp$. We can define our effective perpendicular displacement field by averaging the displacement fields of the metallic nanowires and dielectric host using the metallic fill fraction ( $\rho$ ).

$$D^\perp = \rho D_1^\perp + (1-\rho)D_2^\perp \tag{23}$$

Here $D_1^\perp = \varepsilon_m E_1^\perp$ and $D_2^\perp = \varepsilon_d E_2^\perp$ corresponding to the displacement field of the metallic nanowire and dielectric host respectively. Using these definitions for the displacement field and subbing in our boundary condition (equation 22) into equation 23, we arrive at our final expression for the perpendicular permittivity component for our nanowire metamaterial:

$$\varepsilon_\perp = \rho\varepsilon_m + (1-\rho)\varepsilon_d \tag{24}$$